\documentclass[twocolumn,aps,prl]{revtex4}
\usepackage{amssymb}
\usepackage{graphicx}
\usepackage{graphicx,amsfonts}
\usepackage{bm}

\begin{document}

\title{Evolution of superconducting order in Pr(Os$_{1-x}$Ru$_{x}$)$_{4}$Sb%
$_{12}$}
\author{Elbert E. M. Chia}
\altaffiliation{Los Alamos National Laboratory, Los Alamos, New
Mexico 87545, USA}
\author{M. B. Salamon}
\author{D. Vandervelde}
\affiliation{Department of Physics, University of Illinois at Urbana-Champaign, 1110 W.
Green St., Urbana IL 61801}
\author{D. Kikuchi}
\author{H. Sugawara}
\author{H. Sato}
\affiliation{Department of Physics, Tokyo Metropolitan University, Hachioji, Tokyo
192-0397, Japan}
\date{\today}

\begin{abstract}
We report measurements of the magnetic penetration depth $\lambda$
in single crystals of Pr(Os$_{1-x}$Ru$_{x}$)$_{4}$Sb$_{12}$ down
to 0.1~K. Both $\lambda$ and superfluid density $\rho_{s}$ exhibit
an exponential behavior for the $x$$\geq$0.4 samples, going from
weak ($x$=0.4,0.6), to moderate, coupling ($x$=0.8). For the
$x$$\leq$0.2 samples, both $\lambda$ and $\rho_{s}$ vary as
$T^{2}$ at low temperatures, but $\rho_{s}$ is $s$-wave-like at
intermediate to high temperatures. Our data are consistent with a
three-phase scenario, where a fully-gapped phase at $T_{c1}$
undergoes two transitions: first to an unconventional phase at
$T_{c2}$$\lesssim$$T_{c1}$, then to a nodal low-$T$ phase at
$T_{c3}$$<$$T_{c2}$, for small values of $x$.
\end{abstract}

\maketitle

The recent discovery \cite{Bauer02,Maple02} of the heavy Fermion (HF)
skutterudite superconductor (SC) PrOs$_{4}$Sb$_{12}$ has attracted much
interest due to its differences with the other HFSC. Early work suggested
that the ninefold degenerate $J=4$ Hund's rule multiplet of Pr is split by
the cubic crystal electric field, such that its ground state is a \textit{%
nonmagnetic} $\Gamma _{3}$ doublet, separated from the first excited state $%
\Gamma _{5}$ by $\sim $ 10~K. Hence its HF behavior, and consequently the
origin of its superconductivity, might be attributed to the interaction
between the electric quadrupolar moments of Pr$^{3+}$ and the conduction
electrons \cite{Bauer02}. More recent results appear to rule this mechanism
out, giving strong evidence for a singlet $\Gamma _{1}$ ground state with a $%
\Gamma _{5}$ triplet state at a slightly higher energy
\cite{Aoki02,Goremychkin04}. In this scheme, aspherical Coulomb
scattering \cite{Goremychkin04} and spin-fluctuation scattering
\cite{Hotta04} have been proposed as mechanisms leading to
superconductivity

Surprisingly, replacement of Os by Ru, i.e. in
PrRu$_{4}$Sb$_{12}$, yields a superconductor with
$T_{c}$$\approx$1.25~K \cite{Takeda00} and significantly different
properties. The effective mass of the heavy electrons calculated
from de Haas-van Alphen (dHvA) and specific-heat measurements
\cite{Bauer02,Sugawara02} show that, while PrOs$_{4}$Sb$_{12}$ is
clearly a HF material, PrRu$_{4}$Sb$_{12}$ is at most, a marginal
HF. Various experimental results suggest that these two materials
have different order-parameter symmetry. Firstly, there is no
Hebel-Slichter peak in the
nuclear quadrupole resonance (NQR) data \cite{Kotegawa03} for PrOs$_{4}$Sb$%
_{12}$, while a distinct coherence peak was seen \cite{Yogi03} in the Sb-NQR
1/$T_{1}$ data for PrRu$_{4}$Sb$_{12}$. Secondly, the low-temperature
power-law behavior seen in specific heat \cite{Bauer02} and penetration
depth \cite{Chia03b}, and the angular variation of thermal conductivity \cite%
{Izawa03}, suggest the presence of nodes in the order parameter of PrOs$_{4}$%
Sb$_{12}$. Specifically, Refs.~\onlinecite{Chia03b} and \onlinecite{Izawa03}
reveal the presence of \textit{point} nodes on the Fermi surface (FS). For
PrRu$_{4}$Sb$_{12}$, however, exponential low-temperature behavior was seen
in 1/$T_{1}$ \cite{Yogi03} and penetration depth \cite{Chia04} data. The
latter data were fit with an isotropic zero-temperature gap of magnitude $%
\Delta (0)$=1.9$k_{B}T_{c}$, showing that PrRu$_{4}$Sb$_{12}$ is a
moderate-coupling superconductor. Thirdly, muon spin rotation ($\mu $SR)
experiments on PrOs$_{4}$Sb$_{12}$ reveal the spontaneous appearance of
static internal magnetic fields below \textit{T}$_{c}$, providing evidence
that the superconducting state is a time-reversal-symmetry-breaking (TRSB)
state \cite{Aoki03}. Such experiments have not been performed on PrRu$_{4}$Sb%
$_{12}$.

It is puzzling that the substitution of Ru for Os (same column in the
periodic table) causes PrRu$_{4}$Sb$_{12}$ to differ in so many respects
from PrOs$_{4}$Sb$_{12}$,\ particularly if symmetry of the superconducting
gap varies as we go from Os to Ru. Recently, Frederick \textit{et al.}
performed x-ray powder diffraction, magnetic susceptibility and electrical
resistivity measurements \cite{Frederick04} on single crystals of Pr(Os$%
_{1-x}$Ru$_{x}$)$_{4}$Sb$_{12}$. They found a smooth evolution of
the lattice constant and $T_{c}$ with $x$, \ albeit with a deep
minimum (0.75~K) in $T_{c}$ at $x$=0.6, and an increased splitting
between the ground and excited states of the Pr ion. On the other
hand, one still has to contend with measurements
\cite{Izawa03,Chia03b,Aoki03,Vollmer03} that indicate point-node
gap structure, TRSB and a double superconducting transition
$T_{c2}$$\lesssim$$T_{c}$ \cite{Frederick04} in
PrOs$_{4}$Sb$_{12}$, none of
which are seen for $x$$>$0. We report here a complementary study of Pr(Os$%
_{1-x}$Ru$_{x}$)$_{4}$Sb$_{12}$ using the penetration depth.

A recent paper \cite{Cichorek04} observed an unexpected
enhancement of the lower critical field $H_{c1}(T)$ and the
critical current $I_{c}(T)$ deep in the superconducting state
below $T$$\approx$0.6~K ($T/T_{c}$$\approx$0.3) in
PrOs$_{4}$Sb$_{12}$. They speculate that this reflects a
transition into another superconducting phase that occurs below
$T_{c3}$$\approx$0.6~K, and may explain anomalies in other
measurements, such as the levelling off of Sb-NQR 1/$T_{1}$ below
0.6~K \cite{Yogi03}, the small downturn of
penetration depth below 0.62~K and its deviation from point-node-$T^{2}$%
-behavior above $\sim$0.6~K \cite{Chia03b}.

In this Letter, we present high-precision measurements of the
penetration depth $\lambda (T)$ of
Pr(Os$_{1-x}$Ru$_{x}$)$_{4}$Sb$_{12}$ ($x$=0.1,0.2,0.4,0.6,0.8) at
temperatures down to $\sim $0.1~K using the same experimental
conditions as for PrOs$_{4}$Sb$_{12}$ and PrRu$_{4}$Sb$_{12}$
\cite{Chia03b,Chia04}. For the $x$$\geq$0.4 samples, both $\lambda
$(\textit{T}) and superfluid density $\rho _{s}(T)$ exhibit
exponential behavior at low temperatures, supporting the presence
of an isotropic superconducting gap on the FS. The $\rho _{s}(T)$
data agree with the theoretical curve over the entire temperature
range. The values of $\Delta (0)$ used in the fits suggest an
increase in coupling strength from weak-coupling ($x$=0.4,0.6) to
moderate coupling ($x$=0.8). On the other hand, the $x$$\leq$0.2
samples exhibit a low-$T$ power law, implying the existence of
low-lying excitations. However, the $\rho _{s}$ data fit a
fully-gapped theoretical curve from intermediate temperatures up
to $T_{c}$, but not curves based on a superconducting gap with
line or point nodes. This is consistent with the scenario depicted
by Cichorek \textit{et al.} \cite{Cichorek04}, where for the
$x$$\leq$0.2-samples, the fully-gapped high-$T$ phase undergoes a
transition into a nodal low-$T$ phase below $T_{c3}(x)$. As $x$
increases, the low-$T$ phase is suppressed ($T_{c3}$ decreases)
such that for the $x\geq 0.4$-samples, $T_{c3}$ falls below the
base temperature of our experiment, and we are left with a
fully-gapped phase over our entire experimental temperature range.
Taken together with other data, we suggest that, in addition to
the two phases at $T_{c1}$ and $T_{c2}$, there is a third
superconducting phase at $T_{c3}$ that exhibits point nodes.

The single crystal samples were grown by Sb self-flux method
\cite{Takeda00}. The observation of dHvA effect both in
PrOs$_{4}$Sb$_{12}$ and PrRu$_{4}$Sb$_{12}$ could be an indirect
evidence of high quality of these samples grown in the same
manner. Measurements were performed utilizing a 21-MHz tunnel
diode oscillator \cite{Bonalde2000} with a noise level of 2 parts
in 10$^{9}$ and low drift. The magnitude of the ac field is
estimated to be less than 40~mOe. The sample was mounted, using a
small amount of GE varnish, on a single crystal sapphire rod. The
other end of the rod is thermally connected to the mixing chamber
of an Oxford Kelvinox 25 dilution refrigerator. The sample
temperature is monitored using a calibrated RuO$_{2}$ resistor at
low temperatures (\textit{T}$_{base}$--1.3~K) and a calibrated
Cernox thermometer at higher temperatures (1.2~K--1.8~K).

The deviation $\Delta \lambda (T)$=$\lambda (T)$--$\lambda $(0.1~K) is
proportional to the change in resonant frequency $\Delta $\textit{f}(\textit{%
T}) of the oscillator, with the proportionality factor \textit{G} dependent
on sample and coil geometries. We determine \textit{G} for a pure Al single
crystal by fitting the Al data to extreme nonlocal expressions and then
adjust for relative sample dimensions \cite{Chia03}. Testing this approach
on a single crystal of Pb, we found good agreement with conventional BCS
expressions. The value of \textit{G} obtained this way has an uncertainty of
$\pm$10\% because our samples have a rectangular, rather than square, basal
area \cite{Prozorov2000}.

\begin{figure}[tbp]
\centering \includegraphics[width=7cm,clip]{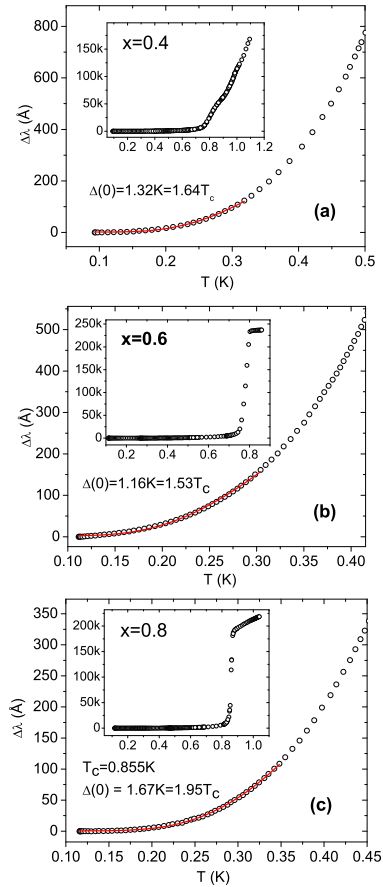}
\caption{($\bigcirc$) Low-temperature dependence of $\Delta \protect\lambda %
(T)$ for (a) $x$=0.4, (b) $x$=0.6, and (c) $x$=0.8. Lines: fits to
BCS low-$T$ expression from $T_{base}$ to 0.4$T_{c}$. The
parameters of the fits are described in the text. Insets show
$\Delta \protect\lambda (T)$ over the full temperature range.}
\label{fig:lambda}
\end{figure}

We first discuss the $x\geq 0.4$ samples. Figure~\ref{fig:lambda}
($\bigcirc $) shows $\Delta \lambda $(\textit{T}) for the three
samples ($x$=0.4,0.6,0.8) as a function of temperature in the
low-temperature region. The insets show $\Delta \lambda
$(\textit{T}) for the entire temperature range. The
onset of the superconducting transitions $T_{c}^{\ast }$ are 0.81~K ($x$%
=0.6) and 0.88~K ($x$=0.8). These values are consistent with those of Ref.~%
\onlinecite{Frederick04}. We could not obtain $T_{c}^{\ast }$ for the $x$%
=0.4 sample as the ac losses were so large that oscillation was lost before $%
T_{c}$ was reached; its large transition width is also consistent with the
ac susceptibility data of Frederick \textit{et al.} \cite{Frederick04},
though the origin is unknown. The values of $T_{c}$, determined from the
point where the experimental superfluid density almost vanishes and fit the
theoretical curves (described later), are 0.8~K ($x$=0.4), 0.76~K ($x$=0.6)
and 0.86~K ($x$=0.8).

For all three samples the data points flatten out below 0.3$T_{c}$, implying
activated behavior in this temperature range. We fit these data to the BCS
low-temperature expression in the clean and local limit, from $T_{base}$ ($%
\sim $0.1~K) to 0.4$T_{c}$, using the expression $\Delta \lambda (T)$%
$\propto $$\sqrt{\pi \Delta (0)/2k_{B}T}\exp (-\Delta
(0)/k_{B}T)$, with the proportionality constant and $\Delta (0)$
as parameters. The best fits (solid lines) are obtained when
$\Delta (0)/k_{B}T_{c}$=1.64 ($x$=0.4), 1.53 ($x$=0.6) and 1.95
($x$=0.8). This implies that the $x$=0.4 and 0.6 samples are
weak-coupling, while the $x$=0.8 sample is a moderate-coupling,
superconductor. The $x$=0.8 result is consistent with that for PrRu$%
_{4}$Sb$_{12}$ ($x$=1).

To extract the superfluid density $\rho _{s}$ from our data, we need to know
$\lambda (0)$. Absent published data on $\lambda (0)$, we assume that it
lies in the vicinity of 344 nm\ (for PrOs$_{4}$Sb$_{12}$) \cite%
{MacLaughlin02} and 290 nm\ (for PrRu$_{4}$Sb$_{12}$)
\cite{Chia04}. We compute $\rho _{s}$ for an isotropic $s$-wave
superconductor in the clean
and local limits using $\rho _{s}=1+2\int_{0}^{\infty }\frac{%
\partial f}{\partial E}d\varepsilon $, where $f$ = [$\exp (E/k_{B}T)+1]^{-1}$
is the Fermi function, and $E=[\varepsilon ^{2}$ + $\Delta
(T)^{2}$]$^{1/2}$ is the quasiparticle energy. The
temperature-dependence of $\Delta (T)$ can be obtained by using
\cite{Gross1986} $\Delta (T)$=$\delta
_{sc}k_{B}T_{c}\tanh \{(\pi /\delta _{sc})\sqrt{(2/3)[(\Delta C)/C][(T_{c}/T)-1]}%
\}$, where $\delta _{sc}$$\equiv $$\Delta (0)/k_{B}T_{c}$ is the
only variable parameter. The specific heat jump $\Delta C/C$ can
be obtained from $\Delta (0)/k_{B}T_{c}$ using strong-coupling
equations \cite{Orlando1979, Kresin1975} .

\begin{table}[tbp]
\begin{tabular}{|c|c|c|c|c|c|c|c|}
\hline
Sample $x$ & 0 & 0.1 & 0.2 & 0.4 & 0.6 & 0.8 & 1.0 \\ \hline
$\Delta (0)/k_{B}T_{c}$ & 2.6 & 1.76 & 1.76 & 1.76 & 1.76 & 1.95 & 1.90 \\
\hline
$\Delta C/C$ & 3.0 & 1.43 & 1.43 & 1.43 & 1.43 & 2.04 & 1.87 \\ \hline
$\lambda (0)$ (nm) & 344 & 320 & 380 & 340 & 380 & 400 & 290 \\ \hline
\end{tabular}%
\\[4.0ex]
\caption{Parameters used to calculate curves in Figs. 2 and 3. Values for $x$%
=0 and $x$=1 are included for comparison.}
\label{table:Table}
\end{table}

\begin{figure}[tbp]
\centering \includegraphics[width=6cm,clip]{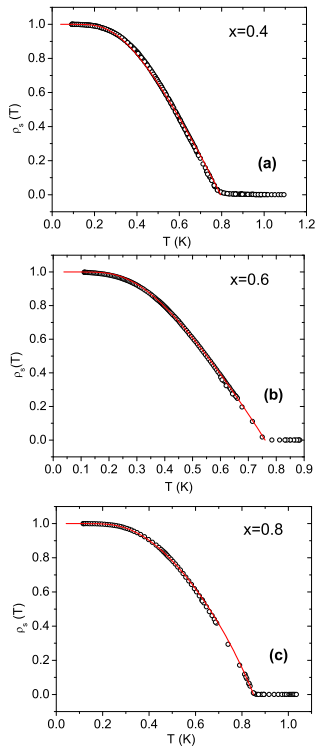}
\caption{($\bigcirc$) Superfluid density $\protect\rho_{s}(T)$ = [$\protect%
\lambda^{2}$(0)/$\protect\lambda^{2}(T)$] calculated from $\Delta \protect%
\lambda (T)$ data in Fig.~\protect\ref{fig:lambda}, for (a) $x$=0.4, (b) $x$%
=0.6, and (c) $x$=0.8. Lines: Theoretical $\protect\rho_{s}(T)$ with
parameters $\Delta (0)/k_{B}T_{c}$ and $\Delta C/\protect\gamma T_{c}$
mentioned in the text.}
\label{fig:rho}
\end{figure}

Fig.~\ref{fig:rho} shows the experimental($\bigcirc $) and calculated (solid
line) values of $\rho _{s}$ as a function of temperature for the $x\geq 0.4$
samples. The theoretical curves fit the data very well using the parameters
shown in Table~\ref{table:Table}. Fitted values for $\lambda (0)$ are
reasonable, considering the uncertainty in obtaining the calibration factor $%
G$.

\begin{figure}[tbp]
\centering \includegraphics[width=8cm,clip]{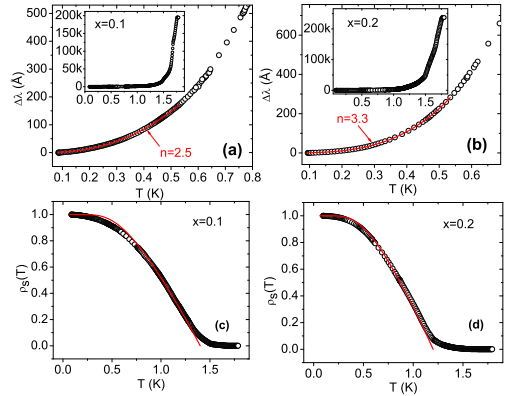}
\caption{($\bigcirc$) Low-temperature $\Delta \protect\lambda (T)$ for (a) $%
x $=0.1 and (b) $x$=0.2. Lines: fits to $\Delta \protect\lambda
(T)$=$A$+$BT^{n}$ from 0.1~K to 0.53~K. Insets show $\Delta
\protect\lambda (T)$
over the full temperature range. ($\bigcirc$) Superfluid density $\protect%
\rho_{s}(T)$ calculated from $\Delta \protect\lambda (T)$ data for (c) $x$%
=0.1 and (d) $x$=0.2. Lines: Theoretical $\protect\rho_{s}(T)$
with weak-coupling parameters. Note the deviation of data from the
theoretical curve at low temperatures is more pronounced for
$x$=0.1 than for $x$=0.2.} \label{fig:Lambda12Rho12}
\end{figure}

We now turn to the $x$$\leq$0.2-samples. Figs.~\ref{fig:Lambda12Rho12}a and %
\ref{fig:Lambda12Rho12}b show $\Delta \lambda (T)$ in the low-temperature
region. The insets show $\Delta \lambda (T)$ for the entire temperature
range. $T_{c}^{\ast}$ is measured to be 1.76~K ($x$=0.1) and 1.77~K ($x$%
=0.2), while $T_{c}$ is 1.4~K ($x$=0.1) and 1.2~K ($x$=0.2). A fit
of the low-temperature data (up to
0.53~K$\approx$0.3$T_{c}^{\ast}$) to a variable power law $\Delta
\lambda (T)$=$A$+$BT^{n}$ yields $n$=2.5 ($x$=0.1) and 3.3
($x$=0.2), indicative of low-lying excitations and incompatible
with an isotropic gap.

\begin{figure}[tbp]
\centering \includegraphics[width=7cm,clip]{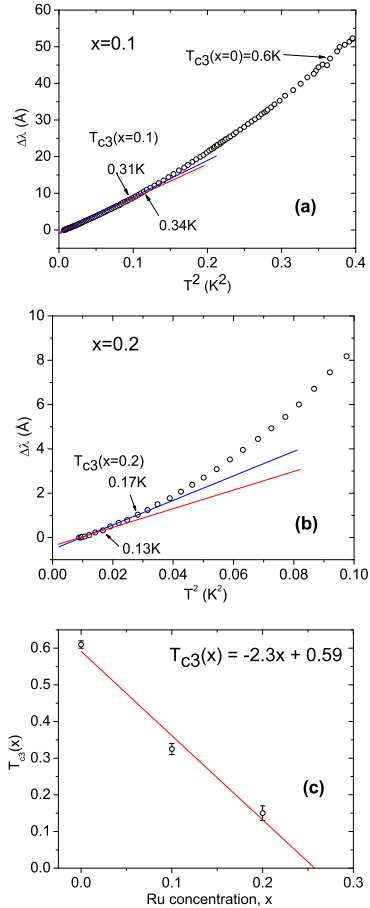}
\caption{($\bigcirc$) Low-temperature $\Delta \protect\lambda (T)$ vs $T^{2}$
for (a) $x$=0.1 and (b) $x$=0.2. The solid lines are visual aids to
determining the range of linear fit. $T_{c3}$ is defined to be the
temperature where $\Delta \protect\lambda (T)$ starts to depart from $T^{2}$%
-behavior. (c) ($\bigcirc$) $T_{c3}(x)$ for $x$=0,0.1,0.2. Line:
Best linear fit to the three data points. Note that the line
extrapolates to zero near $x$=0.26.} \label{fig:lambda12T2Tc3x}
\end{figure}

Figs.~\ref{fig:Lambda12Rho12}c and ~\ref{fig:Lambda12Rho12}d show
the experimental ($\bigcirc $) values of $\rho _{s}(T)$. The solid
lines represents the theoretical curve based on an isotropic
weak-coupling gap as in Table~ \ref{table:Table}. Note that the
data do not agree with the theoretical curve at low temperatures,
but agree from intermediate temperatures up to near $T_{c}$. The
deviation of data from the theoretical curve at low temperatures
is more pronounced for $x$=0.1 than for $x$=0.2. This is
consistent with the scenario depicted by Cichorek \textit{et al.}
\cite{Cichorek04}, where for these low-$x$ samples, the
fully-gapped high-$T$ phase undergoes a transition into a nodal
low-$T$ phase below $T_{c3}(x)$. Our data also agree with the
theory of Hotta \cite{Hotta04}, which predicts that when the
$\Gamma _{1}$-$\Gamma _{5}$ spacing increases (observed as $x$ is
increased from 0 to 1 \cite{Frederick04}, and for $x$=1
\cite{Takeda00}), superconductivity changes from unconventional to
conventional. We assume that this nodal phase is a
\textit{point-node} one, consistent with
Refs.~\onlinecite{Izawa03,Chia03b}, and so $\Delta \lambda
$$\propto$$T^{2}$ in this phase. Consequently, we plot $\Delta
\lambda (T)$ vs $T^{2}$, shown in Fig.~\ref{fig:lambda12T2Tc3x}a
and \ref{fig:lambda12T2Tc3x}b. $T_{c3}(x)$ is determined from the
temperature where the data deviate from linearity, from which we
obtain $T_{c3}$($x$=0.1)$\approx$0.32$\pm$0.02~K and
$T_{c3}$($x$=0.2)=0.15$\pm$0.02~K. Together with
$T_{c3}$($x$=0)$\approx$0.61$\pm$0.01~K deduced in
Ref.~\onlinecite{Cichorek04} and \onlinecite{Chia03b}, we plot
$T_{c3}$ vs $x$ in Fig.~\ref{fig:lambda12T2Tc3x}c. We see that
$T_{c3}$ varies linearly with $x$. Extrapolating the best-fit line
yields $T_{c3}$$\approx$0 when $x$$\approx$0.26. This implies that
the low-$T$ nodal phase disappears, perhaps at a quantum critical
point, when $x\gtrsim 0.3$, i.e. one only sees a fully-gapped
behavior over the whole temperature range, agreeing with our
$x$$\geq$0.4 data sets.

The continuity across the series of the first superconducting
transition, that we label $T_{c1},$ and the BCS-like behavior of
$\rho _{s}$ over much of the $T$-$x$ plane, suggest that
conventional phonon-mediated superconductivity prevails.
Nonetheless, there is ample evidence for a second superconducting
transition at $T_{c2}$ at $x$=0 below which
unconventional superconductivity appears. Specific heat measurements on Pr$%
_{1-y}$La$_{y}$Os$_{4}$Sb$_{12}$ \cite{Rotundu04} showed that the second
superconducting transition at $T_{c2}$ disappears between $y$=0.05 and 0.1,
leaving conventional superconductivity for larger values of $y$. Figs.~%
\ref{fig:lambda}a, ~\ref{fig:Lambda12Rho12}a and
\ref{fig:Lambda12Rho12}b show some changes in curvature in $\Delta
\lambda $ close to $T_{c}^{\ast }$ for the $x$=0.1, 0.2 and 0.4
samples that could be indicative of $T_{c2}$, but which are not
reproducible from sample to sample. As noted in the introductory
paragraph, two mechanisms --- spin-fluctuation and aspherical
Coulomb scattering---have been proposed to explain the
heavy-fermion behavior and superconducting properties of the $x$=0
skutterudite. One possibility is that the spin-fluctuation
mechanism is active at high temperatures where the $\Gamma _{5}$
state is thermally populated on the Os-rich end of the phase
diagram, but is suppressed by decreasing temperature or as Ru
doping increases the $\Gamma _{1}$-$\Gamma _{5}$ splitting.
Aspherical Coulomb scattering may remain important at lower
temperatures and at larger values of $x$. Our data, when
considered together with other data and theory, suggest
\textit{three} different superconducting phases: phonon-driven
(conventional) across the series at the upper transition $T_{c1}$,
but with spin-fluctuation and aspherical Coulomb scattering at the
Os end giving rise to transitions to unconventional phases at
$T_{c2}$ and $T_{c3}$.

In conclusion, we report measurements of the magnetic penetration depth $%
\lambda$ in single crystals of
Pr(Os$_{1-x}$Ru$_{x}$)$_{4}$Sb$_{12}$ down to $\sim$0.1~K. Both
$\lambda$ and superfluid density $\rho_{s}$ exhibit an exponential
behavior for the $x$$\geq$0.4 samples, going from weak-coupling
($x$=0.4,0.6) to moderate-coupling ($x$=0.8). For the $x$$\leq$0.2
samples, both $\lambda$ and $\rho_{s}$ vary as $T^{2}$ at low
temperatures, but $\rho_{s}$ is $s$-wave-like at intermediate to
high temperatures. Our data are consistent with a three-phase
scenario, where a fully-gapped phase at $T_{c1}$ undergoes a
transition to an unconventional phase at
$T_{c2}$$\approx$$T_{c1}$, then to a nodal low-$T$ phase at
$T_{c3}$ for small values of $x$. The $x$-dependence of $T_{c3}$
suggests that the low-$T$ phase disappears near $x$=0.3. 

This material is based upon work supported by the U.S. Department
of Energy, Division of Materials Sciences under Award No.
DEFG02-91ER45439, through the Frederick Seitz Materials Research
Laboratory at the University of Illinois at Urbana-Champaign, and
the Grant-in-Aid for Scientific Research on the Priority Area
``Skutterudites'' (No. 15072206) from MEXT in Japan. Research for
this publication was carried out in the Center for Microanalysis
of Materials, University of Illinois at Urbana-Champaign.
\vspace{-20pt}

\bibliographystyle{prsty}
\bibliography{PrOs4Sb12,CeCoIn5v11}
\bigskip

\end{document}